\documentclass[aps,prd,twocolumn, nofootinbib,showpacs,preprintnumbers]{revtex4}
\usepackage{graphicx}
\usepackage[latin1]{inputenc}

\begin{document}
\preprint{IFF-RCA-010-03-05}
\title{Observing other universe through ringholes and Klein-bottle holes}

\author{Pedro F. Gonz\'{a}lez-D\'{\i}az and Ana Alonso-Serrano}
\affiliation{Colina de
los Chopos, Instituto de F\'{\i}sica Fundamental, \\
Consejo Superior de Investigaciones Cient\'{\i}ficas, Serrano 121,
28006 Madrid, Spain}

\begin{abstract}
It is argued that whereas the Shatskiy single rings produced by
the gravitational inner field of a spherically symmetric wormhole
and the concentric double Einstein rings generated by a toroidal
ringhole could not be used without some uncertainty to identify
the presence of such tunnelings in the universe or the existence
of a parallel universe, the image which the inner gravitational
field of a non orientable Klein-bottle hole tunneling would leave
by lensing a single luminous source is that of a truncated double
spiral, which is a signature that cannot be attributed to any
other single or composite astronomical object in whichever
universe it may be placed. In this report we argue some more
reasons to predict that such a signature would imply the discovery
of one such non orientable tunneling in our or other universe.
After all, a nonorientable Klein-bottle hole  is also a perfectly valid
solution to the Einstein equations and the stuff which would make
it feasible is becoming more and more familiar in cosmology.
\end{abstract}


\maketitle

\section{Introduction}
Considering observable effects from wormholes or ringholes is not
new. More than a decade ago some of such effects were already
predicted by several author [1,2]. In particular, it was shown
that these tunnelings can induce lensing effects from luminous
sources. More precisely, by embedding these tunnelings in
Friedmann space , it was seen [2] that, besides the expected
lensing, at least ringholes were able to induce other potentially
observable effects such as frequency shifting of the emitting
sources, discontinuous change of background temperature, broading
and intensity enhancement of the spectral lines, as well as a
rather dramatic increase in the luminosity of any objects at the
tunnel throat. Moreover, the precise form of the lensing signature
left by wormholes and ringholes has been quite more recently seen
to consist of a single ring and a double concentric ring,
respectively [3,4]. In spite of the interest that all of such
results may have from a pedagogical standpoint, none of these
induced phenomena has by itself any practical usefulness in order
to identify the existence of space-time tunnels in the universe
because there are other observable objects in the universe able to
produce similar effects.

In particular, some hope was raised in using wormholes and
specially ringholes to get a direct evidence of the existence of
other universes to which they could be connected, so traversing
some information from them to ours own. Nevertheless, such a hope
is easily shown to vanish when the following two arguments are
taken into account. First of all, the above alluded uncertainty
that the effects be produced by these tunnelings and not by more
familiar astronomical objects such as galaxies, stars black holes
or quasars.

On the other hand, even though some authors have tried to consider
models where other universes were made observable to us through
collision with our own [5]. Moreover, from the very definition of
universe it follows that provided that there cannot be any
space-time connections between whichever pair of universes, less
yet it is allowed the possibility for any well-defined relation
between the spaces or times of them. It could be argued that if
one allows the connecting wormholes or ringholes to be converted
into time machines with completely unspecified mutual velocities
between the mouths one might create a whole space-time that would
represent two universes. However, such a possibility cannot be
entirely implemented because such a whole space-time would be
orientable against the opposite evidence that it would necessarily
violate orientability. Thus, rather than using orientable
space-time tunnels such as wormhole or ringholes, it appears that
two universes can be mutually tunneled to each other while
preserving their property of still being well-defined universes if
one by instance used non orientable Klein bottle holes converted
into time machines by allowing their mouths to move relative to
one another at completely unspecified speeds [6]. In this paper we
shall consider the effects that such connections would produce and
the possibility to using them to check the existence of universes
other than ours.

One of the most important revolutions in cosmology is taking place
right now, that of the so called multiverse [7]. In order to
convert this revolution in more than just a speculative idea,
providing it with a physical content, one should try to consider
alternate ways which would for example include the effects that
time machines derived from nonorientable Klein bottle holes [6]
may have on the luminous sources placed beyond the limit of our
universe. Something that any of the existing multiverse theories
is very needed of. In fact, Linde ideas coming about string
theories and alike [8] or other's ideas [9] may make the notion of
a multiverse more plausible, but they do not prove that other
universes are really out there. The staggering challenge is to
think of a conceivable experiment or observation confined to our
own universe based on looking for some footprints left by
nonorientable tunnelings connecting our universe to other
universes. General principles of physics cast in fact serious
doubts on whether it make sense to talk about other universes if
they can never be detected.

Rees, an early supporter of multiverse idea, agrees [10] that it
may never be possible to observe other universes directly, but he
argues that scientists may still be able to make a convincing case
for their existence. To do that, he says, physicists will need a
theory of the multiverse that makes new but testable predictions
about properties of our own universe. If, similarly to as current
observations have confirmed big bang as a well established model,
new experiments coming perhaps from Large Hadron Collider or the
Planck satellite space mission, indirectly confirmed such a theory
predictions about the universe we can see, Rees believes, they
would also make a strong case for the reality of those we cannot.
String theory is still very much a work in progress, but it could
form the basis for the sort of theory that Rees is calling for.

However, the very essentials of quantum theory show some great,
almost insurmountable odds against the Rees philosophy, specially
if one adheres to the quantum-cosmological ideas that support the
principle according to which physical reality should be directly
observable or it vanishes into nothing. In this way, it appears a
casus of full necessity to explore the existing ways that may lead
us to directly observe a property or characteristic of a universe
other than ours, such as the main objective of the present paper
is aiming at exploring the possible existence of other universes
through a search for the lensing signature of orientable ringholes
and non orientable Klein bottle holes connecting us to other
universe.

\section{Wormhole signature}

It was first noted by Shatskiy [3] that wormholes, which are
usually disguised as black holes, can be made observable and
recognizable in terms of bright, glowing rings originating from
the necessary flaring out of the embedding surface around their
throat which is produced by the presence of the so called phantom
stuff [11]. The really most devastating argument against the
proposed wormhole distinguishable character of the Shatskiy rings
is that, even if exotic matter does exist, other many objects are
able to create a similar lensing light signature [12]. In
particular, while the orientability of the wormhole manifold makes
it impossible using these solutions for observing any thing from a
universe other than ours, it is hard to see how the resulting
lensing rings could be differentiated from the astronomical
blueprint left by negative energy stars and, mainly, from all
those massive astronomical objects, such as galaxies, black holes
or quasars, whose gravitational lensing effects appear as the so
called Einstein rings [11].

The actual problems are with the symmetry and the orientability of
the throat. Wormholes are orientable manifolds which are
characterized by a spherically symmetric throat and, therefore,
there will be a diverging lensing effect undergone by any bundle
of light rays coming from a luminous source placed behind the
furthest wormhole mouth which would necessarily manifest to
observers on Earth as single perfectly circular rings, such as it
was indicated by Shatskiy [3]. This pattern could well be
misinterpreted as being originated from a star or other massive
astronomical object necessarily placed in our own universe,
instead of a wormhole, with a radius quite smaller than that for
that wormhole throat radius.

\begin{figure}
\includegraphics[width=.9\columnwidth]{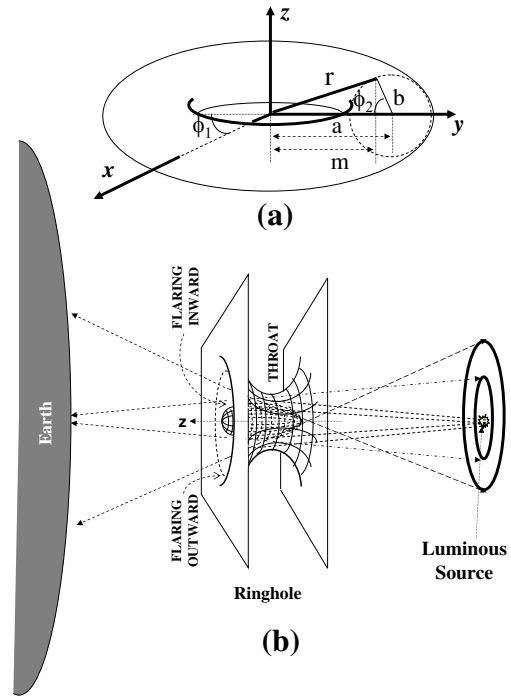}
\caption{\label{fig:epsart} Gravitational lensing effect produced
by a ringhole from a single luminous source. (a) Parameters
defining the toroidal ringhole throat in terms of which metric (1)
is defined. (b) Rays passing near the outer and inner surfaces
respectively flare outward and inward, leading to a image from a
luminous point placed behind the ringhole which is made of two
concentric bright rings. The relative mutual positions of these
rings would depend on the distance between the ringhole and the
luminous source. If that distance is small enough then the larger
outer ring comes from the flaring inward surface, and conversely,
if the distance source-ringhole is increased then the outer ring
comes from the outward surface, the larger that distance the
greater the difference between the two bright rings.}
\end{figure}

\section{Ringhole signature}

An inner tunneling symmetry which would give rise to a more
distinguishable lensing pattern is that of a ringhole [13], that
is, an orientable space-time tunnel whose throat has the toroidal
symmetry, instead of the spherical symmetry. Using the set of
geometrical parameters specified in the upper part of Fig. (1) we
can derive the metric for a ringhole to be [13]
\begin{equation}
ds^2= -C_2 r^2 dt^2 +b^2\left[1+\frac{C_1
a^2\sin^2\varphi_2}{r^6\left(1-
\frac{A^2}{r^4}\right)}\right]d\varphi_2^2 + m^2 d\varphi_1^2
\end{equation}
where
\begin{equation}
A=a^2-b^2,\;\; m=a-b\cos\varphi_2 ,\;\; r=\sqrt{a^2+b^2-2ab\cos\varphi_2},
\end{equation}
with $C_1$ and $C_2$ arbitrary integration constants, and $a$ and
$b$ the radius of the circumference generated by the circular axis
of the torus and that of a torus section, respectively, with
$a>b$. Metric (1) is defined for $0\leq t\leq\infty$, $a-b\leq
r\leq a+b$ and the angles (see Fig. 1 (a)) $0\leq \varphi_1$,
$\varphi_2 \leq 2\pi$.

A thorough analysis [4,13] leads then us to deduce that one would
expect lensing effects to occur at or near the ringhole throat,
that is to say, the mouths would act like a diverging lens for
world lines along $2\pi - \varphi_2^c
>\varphi_2 >\varphi_2^c$, and like a converging lens for world
lines along $- \varphi_2^c <\varphi_2 <\varphi_2^c$. No lensing
actions would therefore take place at $\varphi_2=\varphi_2^c$ and
$\varphi_2=2\pi - \varphi_2^c$. In fact, in the case of ringholes,
instead of producing just a single flaring outward for light rays
passing through the wormhole throat, this multiply connected
topology, in addition to that flaring outward (diverging) effect,
also produces a flaring inward (converging) effect$^{13}$ on the
light rays that pass through its throat, in such a way that an
observer on Earth would interpret light passing through the
ringhole throat from a single luminous source as coming from two
bright, glowing concentric rings, which form up the distinctive
peculiar pattern from ringholes (See Fig. 1 (b)). That pattern
cannot be generated by any other possible disturbing astronomical
object other than a very implausible set of three luminous massive
objects (let us say galaxies) which must be so perfectly aligned
along the sigh line that its occurrence becomes extremely
unlikely.

It is readily inferred from Fig. 1 (b) that, for a reasonably
large ringhole sufficiently far from the luminous source, the
inner bright ring would correspond to the flaring inward
(converging) surfaces. If we keep the ringhole size invariant and
the distance between the ringhole and the luminous source is
decreased drastically, then the inner bright ring would turn to be
produced by the flaring outward (diverging) surface.

Such a ringhole signature may have been already observed though it
has been so far attributed to the combined effect of two Einstein
rings originated from the above-considered to be extremely
unlikely superprecise alignment of three galaxies. In fact, at the
beginning of 2008 The NASA/ESA Hubble Space Telescope revealed
[14] a never-before-seen phenomenon in space: a pair of glowing
rings, one nestled inside the other like a bull's-eye pattern.
This double-ring pattern was interpreted as a double Einstein ring
being caused by the complex bending of light from two distant
galaxies strung directly behind a foreground massive galaxy, like
three beads on a string along the line of sight, simply because at
the time there were no other available interpretations for what
was being observed. Being more than just a novelty, this very rare
phenomenon found with the Hubble Space Telescope could, moreover,
eventually offer insight into dark matter, dark energy, the nature
of distant galaxies, and even the curvature of the Universe.

As previously stated, for that interpretation to be feasible, the
massive foreground galaxy had to be almost perfectly aligned in
the sky with two background galaxies at different distances to
justify the finding. The foreground galaxy is 3 billion
light-years away. Now, in order to justify the ratio between the
two ring radii, the inner ring and outer ring would be comprised
of multiple images of two galaxies  at a distance of some 6
billion and approximately 11 billion light-years.

However, the odds of observing the required extremely precise
alignment of the three galaxies are so small (an estimated 1 in
10,000) that even some of the discoverers of that astronomical
phenomena said that they had 'hit the jackpot' with the discovery.
At the time, the authors of Ref. 14 had no alternative other than
accepting that quite improbable interpretation of the result.
Nevertheless, having we uncovered that such concentric rings may
well be also interpreted as the blueprint of the presence of a
ringhole in the direction in space where the double bright ring
system was discovered, we may also adopt the latter interpretation
in terms of a ringhole as an alternate possible explanation for
that phenomenon, taking now the luminous sources at redshifts
corresponding to 3 and 6 billion light-years as measuring the
positions of the two ringhole mouths on the sky (provided the
mouths are surrounded by some sufficiently large quantities of
luminous material), and their respective luminosities as stemming
from the respective light deflections along the angle $\varphi_2$
caused by the combined effect of the size of the throat radius and
the relative distance between the two mouths. Even in the absence
of sufficiently high matter densities around the two ringhole
mouths in a still large enough tunneling, since the two rings have
the same spectra in the case of a ringhole, it could well be that
the unseen second source might be too faint so that these spectra
actually be the same which corresponds to matter placed anywhere
along the ringhole tunneling. The feature that the surface
brightness of the two rings would be different may also be
justified by the above-suggested different distribution of
ordinary and exotic matter leading to distinct absorption and
dispersion effects of the incoming light along the two horizon
separated $\varphi_2$-angular regions around the throat. In any
event, because of the orientability of the single ringhole
manifold or the composite manifold formed by the aligned galaxies,
which is required to get the observed concentric double ring
bright image, the luminous source originating the lensing effect
can never be placed in a universe other than ours, making in this
way a ringhole a completely useless tool to check the possible
existence of other universes.

\section{Klein bottle signature}

The use of inter-universe tunnelings in order to check the
existence of other universes should actually require choosing
spacetime holes which be (i) non orientable and (ii) convertible
into a time machine with completely unspecified relative speed
between its mouths. One such space-time construct has in fact been
already studied, in the shape of what is dubbed a Klein bottle
hole [6]. In this case, non orientability is guaranteed by the
existence of a throat with the topology of a Klein bottle and it
was also shown that this space-time can be stable to vacuum
fluctuations and is also convertible into a time machine with
fully arbitrary inter mouths speed. Employing the set of
geometrical parameters specified in Fig. 2 and in the lower part
of Fig. 3 we can derive the space-time metric for one of such
Klein bottle holes to be [6]
\begin{eqnarray}
&&ds^2= -e^{2\Phi}dt^2+\theta(2\pi-\varphi_1)\left(\frac{dr_1^2}{1-K(b_1)/b_1} +d\Omega_1^2\right)\nonumber\\
&&+\theta(\varphi_1-2\pi)\left(\frac{dr_2^2}{1-K(b_2)/b_2}+d\Omega_2^2\right) ,
\end{eqnarray}
where the $\theta(x)$'s are the step Heaviside function [15], with
$\theta(x)=1$ for $x>0$ and $\theta(x)=0$ for $x<0$,
$d\Omega_i^2$'s for $0\leq\varphi_1\leq 2\pi$ is given by
\begin{eqnarray}
&&d\Omega_1^2=\left\{m_1^2+
\frac{1}{4}\left[M_1(a_1-C_1)+N_1(b_1-D_1)\right]\right\}d\varphi_1^2\nonumber\\
&&b_1^2d\varphi_2^2-b_1\sqrt{(a_1-C_1)(A_1-a_1)}\sin\varphi_2d\varphi_1
d\varphi_2  ,
\end{eqnarray}
in which
\begin{equation}
M_1=A_1-a_1-(B_1-b_1)\cos\varphi_2
\end{equation}
\begin{equation}
N_1=B_1-b_1-(A_1-a_1)\cos\varphi_2
\end{equation}
\begin{equation}
m_i=a_i-b_i\cos\varphi_2 ,\;\; i=1,2
\end{equation}
Now, for $2\pi\leq\varphi_1\leq 3\pi$
\begin{eqnarray}
&&d\Omega_1^2=\left\{m_2^2+
\frac{1}{4}\left[M_2(a_2-A_2)+N_2(b_2-B_2)\right]\right\}d\varphi_2^2\nonumber\\
&&b_2^2d\varphi_2^2-b_2\sqrt{(a_2-A_2)(C_2-a_2)}\sin\varphi_2
d\varphi_1 d\varphi_2  ,
\end{eqnarray}
where in this case
\begin{equation}
M_2=C_2-a_2-(D_2-b_2)\cos\varphi_2
\end{equation}
\begin{equation}
N_2=D_2-b_2-(C_2-a_2)\cos\varphi_2 .
\end{equation}

Finally we have
\begin{equation}
r_1 = \sqrt{a_1^2+b_1^2-2a_1 b_1\cos\varphi_2}
\end{equation}
\begin{equation}
r_2 = \sqrt{a_2^2+b_2^2+2a_2 b_2\cos\varphi_2},
\end{equation}
where we have extended the range of the angular coordinate
$\varphi_1$ to also encompass values continuously running from
$2\pi$ to $3\pi$ and $A_i$, $B_i$, $C_i$ and $D_i$, $i=1,2$, are
given sets of adjustable parameters which are arbitrary unless for
the conditions $A_1 > C_1$, $B_1
>D_1$, $A_1
>B_1$ and $C_1 >C_1$ for the angular interval $0\leq\varphi_1\leq
2\pi$ ,whereas for $2\pi\leq\varphi_1\leq 3\pi$ we must have $C_2
>A_2$, $D_2 >B_2$, $C_2 >D_2$ and $A_2 >B_2$, with $D_2=B_!$,
$B_2=D_1$ and $A_1-C_1 = A_2+C_2$, with $A_1-C_1 >2A_2$. $a_i$ and
$b_i$ are the radius of the circumference generated by the
circular axis of the Klein bottle torus and that of a Klein bottle
section, respectively, with $a_i>b_i$. Metric (3) is defined for
$0\leq t\leq\infty$, $a_i-b_i\leq r_i\leq a_i+b_i$ and the angles
(see Figs. 2 and 3) $0\leq \varphi_1$, $\varphi_2 \leq 2\pi$.

\begin{figure}
\includegraphics[width=.9\columnwidth]{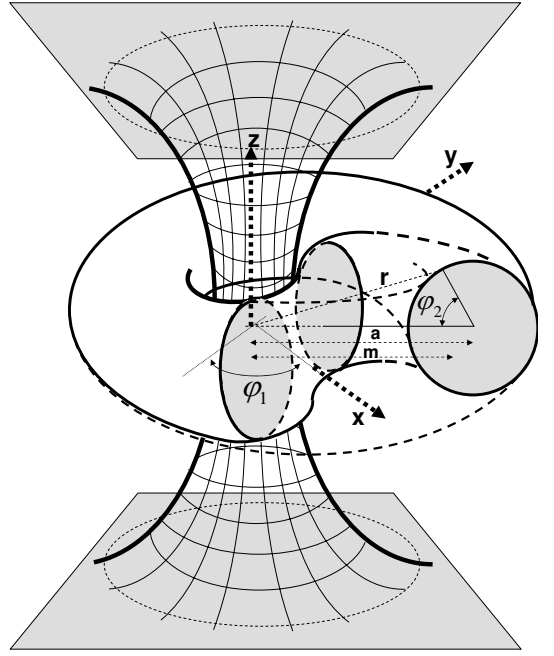}
\caption{\label{fig:epsart} Pictorial representation of a Klein
bottle hole showing the non orientable topology of its throat and
some of the parameters in terms of which metric is defined (see
also Fig. 3).}
\end{figure}

In order to check the properties of a Klein bottle hole as a lens,
we now write the static space-time metric of a single, traversible
Klein bottle hole in the form
\begin{eqnarray}
&&ds^2 =-dt^2
+\theta(2\pi-\varphi_1)\left[\left(\frac{n_1(\ell_1)}{r_1(\ell_1)}\right)^2
d\ell_1^2 +d\Omega_1^2(\ell_1)\right]\nonumber\\
&&+\theta(\varphi_1
-2\pi)\left[\left(\frac{n_2(\ell_2)}{r_2(\ell_2)}\right)^2
d\ell_2^2 +d\Omega_2^2(\ell_2)\right] ,
\end{eqnarray}
where $-\infty <t < +\infty$, with $-\infty <\ell_i < +\infty$,
and $i=1,2$. Here, $b_i$ is replaced for $\sqrt{\ell_i^2+b_0^2}$,
$\ell_i$ being the proper radial distance of each transversal
section of the Klein bottle on the respective $\varphi_1$ interval
for $i$, and
\begin{equation}
m_{i}(\ell_i)=a_i-\left(\ell_i^2
+b_{0i}^2\right)^{1/2}\cos\varphi_2\nonumber
\end{equation}
\begin{equation}
n_{i}(\ell_i)=\left(\ell_i^2
+b_{0i}^2\right)^{1/2}-a_i\cos\varphi_2 ,
\end{equation}
\begin{equation}
r_{i}(\ell_i)=\sqrt{a_i^2 +\ell_i^2 +b_{0i}^2
+2(-1)^i\left(\ell_i^2 +b_{0i}^2\right)^{1/2}a_i\cos\varphi_2} ,
\end{equation}
in which $b_{0i}$ is as given from
$b_1=(B_1-D_1)\cos^2(\varphi_1/4)+D_1$ and
$b_2=(D_2-B_2)\sin^2(\varphi_1/2)+B_2$ for constant parameter
adjusted to be the radius of the throat of the Klein bottle hole
at $\ell_i=0$. As $\ell_i$ increases from $-\infty$ to 0, $b_i$
decreases monotonously from $+\infty$ to its minimum value
$b_{0i}$ at the throat radius, and as $\ell_i$ increases onward to
$+\infty$, $b_i$ increases monotonously to $+\infty$ again. Now,
for metrics (3) and (13) to describe a Klein bottle hole we must
embed it in a three-dimensional Euclidean space with cylindrical
coordinates at fixed time $t$ [6] whose metric can be written as
\begin{equation}
ds^2=dz^2 +dr^2 +r^2 d\phi^2
=\left[1+\left(\frac{dz}{dr}\right)^2\right]dr^2 +r^2 d\phi^2 ,
\end{equation}
However, since $r_i$ and $\varphi_1$ are no longer independent to
each other, one can always convert metric (3) into a line element
which is embeddible in the cylindrical space (17). That conversion
can be made by first obtaining the expression for the variation of
$r_i$ with respect to $\varphi_1$, that is
\begin{equation}
\frac{dr_i}{d\varphi_1}=Q(i)=-\frac{\left[m_i(A_i-C_i)
+n_i(B_i-D_i)\right]\sin(i\varphi_1/2)}{2ir_i} .
\end{equation}
From which we get
\begin{eqnarray}
&&ds^2 =-e^{2\Phi}dt^2
+\theta(2\pi-\phi_1)\left(\frac{c(1)dr_1^2}{1-b_{01}^2/b_1^2}\right.\nonumber\\
&&\left.+d(1)Q(1)dr_1 d\varphi_1 +d\Omega_1^2\right) +
\theta(\varphi_1-2\pi)\nonumber\\
&&\times\left(\frac{c(2)dr_2^2}{1-b_{02}^2}+ d(2)Q(2)dr_2
d\varphi_1 +d\Omega_2^2\right) ,
\end{eqnarray}
with $c(i)+d(i)=1$.

Taking now $dz=(dz/dr_1)dr_1 +(dz/d\varphi_1)d\varphi_1$, for
$\varphi_1\leq 2\pi$ and $dz=(dz/dr_2)dr_2
+(dz/d\varphi_1)d\varphi_1$, for $\varphi_1 > 2\pi$ we can derive
for any allowed value of $\varphi_2$,
\begin{equation}
c(i)=1+2\left(1-\frac{b_{0i}^2}{b_i^2}\right)
-2\sqrt{1-\frac{b_{0i}^2}{b_i^2}} .
\end{equation}
Hence it follows that the metric for the non orientable Klein
bottle hole which is embeddible in flat space is described by Eq.
(19) with $c(i)$ as given by (20) and $d=1-c(i)$. Using these
coefficients, metric (17) will be the same as metric (19) for
constant values of $\varphi_2$ if we identify the coordinates $r$,
$\phi$ of the embedding space with either coordinates $r_1$,
$\phi_1$, for $\varphi_1\leq 2\pi$, or the coordinates $r_2$,
$\varphi_1$, for $\varphi_1 > 2\pi$, and if we require the
function $z$ to satisfy
\begin{equation}
dz/dr_i=1+\left(1-\frac{b_{0i}^2}{b_i^2}\right)^{-1}-
\left(1-\frac{b_{0i}^2}{b_i^2}\right)^{-1/2},
\end{equation}
for any value of $\varphi_1$,
\begin{equation}
\frac{dz}{d\varphi_1}=\frac{1}{2}\sqrt{\left[R(\varphi_2)_1
-r_1\right]\left[r_1-\rho(\phi_2)_1\right]},
\end{equation}
for $\varphi_1\leq 2\pi$, and
\begin{equation}
\frac{dz}{d\varphi_1}=\sqrt{\left[R(\varphi_2)_2
-r_2\right]\left[r_2-\rho(\phi_2)_2\right]},
\end{equation}
for $\varphi_1 > 2\pi$, where
\begin{equation}
R(\varphi_2)_1= A_1-B_1 \cos\varphi_2,\;\; \rho(\varphi_2)_1 = C_1
-D_1\cos\varphi_2
\end{equation}
and
\begin{equation}
R(\varphi_2)_2= C_2-D_2 \cos\varphi_2,\;\; \rho(\varphi_2)_2 = A_2
-B_2 \cos\varphi_2 .
\end{equation}
From these expressions and the requirement that non orientable
Klein bottle holes be connectible to asymptotically flat
space-time, one can deduce how the embeddible surfaces would flare
at or around the hole throat. Thus, from Eq. (21), one obtains
\begin{eqnarray}
&&\frac{d^2 r}{dz^2} = \frac{b_{0i}^2 r_i}{b_i^3
n_i}\left(\frac{1}{\sqrt{1-b_{0i}^2/b_i^2}}-1\right)\nonumber\\
&&\times \left(1+\frac{1}{1-b_{0i}^2/b_i^2}
-\frac{2}{\sqrt{1-b_{0i}^2/b_i^2}}\right)^{-7/2} ,
\end{eqnarray}
which is positive for $2\pi-\varphi_2^c >\varphi_2 >\varphi_2^c
=\arctan(b_i/a_i)$ and negative for $-\varphi_2^c <\varphi_2
<\varphi_2^c$. Thus, exactly to as it happens in the case of
toroidal ringholes$^{4,13}$, the embedding surface flares outward
for $d^2 r/dr^2 >0$ and flares inward $d^2 r/dr^2 <0$. It is for
this reason that a Klein bottle hole would generally behave like a
diverging lens for $2\pi-\varphi_2^c >\varphi_2 >\varphi_2^c$, and
like a converging lens for $2\varphi_2^c <\varphi_2 <\varphi_2^c$,
even though, unlike in the ringhole case, these behaviors will
also depend on the value of the angle $\varphi_1$ due to the non
orientable character of the Klein bottle hole space-time, such as
we shall show in some detail in what follows.

In order to investigate how the embedding surface flares at or
around the throat as the angle $\varphi_1$ is varied, so making
the manifold non orientable, we distinguish two cases. The first
case corresponds to condition (22), from which we can derive
\begin{widetext}
\begin{equation}
\frac{d^2\varphi_1}{dz^2}= \frac{\left[-2r_1+R(\varphi_2)_1+
\rho(\varphi_2)_1\right]\left[R(\varphi_2)_1-
\rho(\varphi_2)_1\right]\sin(\varphi_1/2)}{2\left\{\left[R(\varphi_2)_1-
r_1\right]\left[r_1 -\rho(\varphi_2)_1\right]\right\}^2} .
\end{equation}
\end{widetext}
Now, since $a_i>b_i$ for $0\leq\varphi_1\leq 2\pi$ one obtains
that this expression vanishes at $\varphi_1 =\pi$ and becomes
negative for  $\varphi_1 < \pi$, on which values the embedding
surface flares toward larger values of the radius $b_1$, and
negative for $\varphi_1 >\pi$ on whose values the embedding
surface flares toward smaller values of $b_1$.

The second case corresponds to condition (23) for which we get
\begin{widetext}
\begin{equation}
\frac{d^2\varphi_1}{dz^2}= \frac{\left[-2r_2+R(\varphi_2)_2+
\rho(\varphi_2)_2\right]\left[R(\varphi_2)_2-
\rho(\varphi_2)_2\right]\sin(\varphi_1)}{4\left\{\left[R(\varphi_2)_2-
r_2\right]\left[r_2 -\rho(\varphi_2)_2\right]\right\}^2} .
\end{equation}
\end{widetext}
The critical value of $\varphi_1$ becomes then $\varphi_1=5\pi/2$.
For $\varphi_1<5\pi/2$, Eq. (28) becomes negative so that the
embedding surface flares toward smaller values of $b_2$, while it
becomes positive for $\varphi_1>5\pi/2$, for which the embedding
surface flares toward larger values of $b_2$.

The above analysis leads us to expect lensing effects to occur on
the mouths of the nonorientable Klein bottlehole with respect to a
bundle of light rays, at or near the throat, coming from the
distribution of negative and positive values for the energy
density [6]; that is, the mouths would act like a diverging lens
for world lines along the values of the coordinates, at or near
the throat, which correspond to negative energy density, and like
a converging lens for world lines passing through regions with
positive energy density. In order to confirm with full accuracy
which regions around the throat behave like a converging or
diverging lens, one must consider the null-ray propagation
governed by the integral of the stress-energy tensor. From the
Einstein equations [6] it can be obtained that the mouths focus or
defocus a bundle of rays, depending on the sign of the integral
[6]
\begin{widetext}
\begin{eqnarray}
&&I=\int_{\ell_i^1}^{\infty}d\ell_i e^{\Phi}\left(\rho
c^2-\sigma\right)=\int_{\ell_i^1}^{\infty}d\ell_i
e^{\Phi}\left(\frac{n_1}{r_1}\right)^3\left(T_0^0
-T_1^1\right)=\int_{\ell_i^1}^{\infty}d\ell_i e^{\Phi}\frac{c^4
b_{01}^2 n_1^2}{16\pi Gb_1^3 r_1^2}\times\nonumber\\
&&\left\{\frac{\frac{8m_1}{a_1}-\left(\frac{A_1 -C_1}{a_1}
+\frac{B_1 -D_1}{b_1}\right)\cos^2\left(\frac{\varphi_1}{4}\right)
+\frac{m_1^{(0)}}{m_1} +\frac{n_1^{(0)}}{n_1}-2}{4\left\{m_1^2
+\frac{1}{4}\left[M_1(A_1-C_1)+N_1(B_1-D_1)\right]\cos^2\left(\frac{\varphi_1}{4}\right)\right\}}
+\frac{2\left(1+\sin\varphi_2\right)}{n_1 b_1}
-\frac{4cos\left(\frac{\varphi_1}{2}\right)\sin\varphi_2}{(A_1
-C_1)m_1 \sin^2\left(\frac{\varphi_1}{2}\right)}\right\}
\end{eqnarray}
\end{widetext}
In what follows we shall restrict ourselves to consider the case
$\Phi=0$ and the still realistic situation where $\varphi_2=\pi/2$
and $a_1 >> b_1$. We obtain then
\begin{eqnarray}
&&I=\frac{c^4}{16\pi
G}\left[\xi_1(\varphi_1)\left(\frac{\pi}{2}-\arctan\left(\frac{\ell_{1}^1}{b_{01}}\right)\right)\right.\nonumber\\
&&\left.+\xi_2(\varphi_1)\left(1-\frac{\ell_1^1}{\sqrt{(\ell_1^1)^2
+b_{01}^2}}\right)\right]
\end{eqnarray}
where
\begin{equation}
\xi_1(\varphi_1)=\frac{(B_1-D_1)\cos^2\left(\frac{\varphi_1}{4}\right)+B_1}{a_1^2\left[4a_1^2
+(A_1 -a_1)(a_1-C_1)\right]}
\end{equation}
\begin{equation}
\xi_2(\varphi_1)=\frac{4}{a_1^2} .
\end{equation}
We can conclude in this way that the integral $I$ is always
definite positive, irrespective of the value of the angle
$\varphi_1$. It follows that the dependence of $I$ on $\varphi_1$
will only contribute the strength of the action of the two
different gravitational lenses that can be distinguished at and
around the Klein-bottle hole throat, so just quantitatively
modifying the behavior governed by the angle $\varphi_2$.

Inspection on the above equations leads to the final result which
is twofold. On the one hand, we can derive that the surface
gravity $\kappa$ is definite positive for $2\pi-\varphi_2^c
>\varphi_2 >\varphi_2^c$ and definite negative for $-\varphi_2^c
<\varphi_2 <\varphi_2^c$, and hence, since generally we have
$T=-\kappa|_{b_i=b_{0i}}/2\pi$, that the Klein bottle hole emits
thermal (phantom) radiation at negative temperature from the first
of these regions and thermal radiation at positive temperature
from the second region [16]. On the other hand, a ready
calculation leads us to obtain (see Fig. 3) that the signature
that a Klein bottle hole would in any event leave from a luminous
source placed beyond it for an observer on earth is that of two
concentric truncated double spiral.

\begin{figure}
\includegraphics[width=.9\columnwidth]{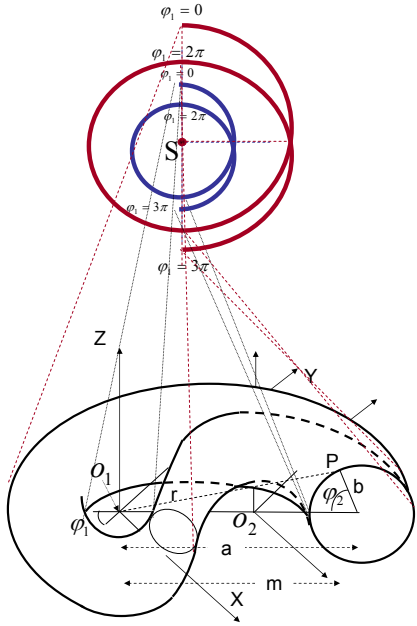}
\caption{\label{fig:epsart} Gravitational lensing effect produced
by a Klein bottle hole from a single luminous source. On the lower
part of the figure the parameters defining the Klein bottle hole
throat in terms of which metric (3) is defined. The rays passing
near the outer and inner surfaces respectively flare outward and
inward, leading to a image from a luminous point placed behind the
Klein bottle hole which is made of two concentric truncated bright
spirals.}
\end{figure}

In this case, besides valuable information on dark matter, dark
energy and universe curvature or the very existence of Klein
bottle holes in our universe, what could eventually be most
astonishing in its implications would be an unprecedented insight
into some of the contents of other universes linked to ours by
means of these Klein bottle holes. After all, a Klein bottle hole
is a perfectly valid solution to the Einstein equations for
stress-energy tensor containing a given proportion of exotic stuff
- possibly phantom energy- which is becoming more and more
familiar in the full context of current cosmology [17]. The
potentially attainable insight from such an interpretation is
twofold. On the one hand, we would get a direct evidence for the
existence of Klein bottle holes and, by the way, possibly of
wormholes and ringholes, and on the other hand, we could have
found a way to open the door to observe a parallel universe, and
hence got a first direct evidence for the existence of the
multiverse scenario.

There is an observation which may in principle distinguish a
static Klein bottle hole staying within our own universe and
having its two mouths at rest with respect to each another, from a
Klein bottle hole that connect our universe to a parallel universe
or, in general, to other universe of a multiverse scenario. In the
latter case since there is no common space-time for the two
universes (parallel or not), the two mouths should necessarily be
in perpetual quasi periodic relative random motion with completely
unspecified speed. This would make the time and space for the two
universes at all independent because the relative motion of the
two mouths converts the Klein bottle hole in a time machine that
contains completely arbitrary closed timelike curves. In the case
of the inner static Klein bottle hole, if the luminous source is
kept motionless and the Klein bottle hole does not behave like a
time machine, the double truncated spirals would be well resolved
and defined on the pattern. However, if the positions of the two
mouths continuously vary relative to each other in a random though
quasi periodic way then the width of each of the truncated spirals
would be stretched out and their resolution spoiled and clearly
blurred due to the continuous and completely arbitrary changes of
distance between the two mouths, thus leading to a glowing
background around the spirals of Fig. 3, showing just a maximum of
intensity at the average relative position of the mouths, provided
the relative motion keep a sufficiently high degree of
periodicity. In the latter case, the metric of the Klein bottle
hole would change to be given by a line element that describes
arbitrary time travel induced by a nearly periodic relative motion
between the two mouths. Using arguments similar to those employed
in Refs. [4] and [5] we finally get for that line element
\begin{eqnarray}
&&ds^2=-e^{2\Phi}dt^2
+\theta(2\pi-\varphi_1)\left(\left\{-[1+\bar{g}F(\ell_1)\ell_1\sin\varphi_2\right]^2\right\}.\nonumber\\
&&\left.\times e^{2\Phi}dt^2 +c(1)d\ell_1^2+d(1)Q(1)dr_1
d\varphi_1
+d\Omega_1^2\right)\nonumber\\
&&+\theta(\varphi_1 -\pi)\left(\left\{-[1+\bar{g}F(\ell_2)\ell_2\sin\varphi_2\right]^2+1\right\}\nonumber\\
&&\left.\times e^{2\Phi}dt^2 +d(2)Q(2)dr_2 d\varphi_1
+d\Omega_2^2\right),
\end{eqnarray}
where $\bar{g}=\bar{\gamma}^2\frac{d\bar{v}}{dt}$ is the average
acceleration of the moving mouth, with $\bar{v}$ the corresponding
average velocity, and $\bar{\gamma}=1/\sqrt{1-\bar{v}^2}$ is the
average on the fuzzy relativistic factor; finally $F(\ell)$ is a
form factor that vanishes in the half of the Klein bottle hole
which is assumed to be kept motionless, and rises up on average
from 0 to 1 as one moves along the direction of the moving mouth.
We must finally point out that any ringhole which is a time
machine even within our universe will also show an defocused
double truncated spiral pattern though not so blurred perhaps as
that corresponding to an inter-universe ringhole.

\section{Further comments}

We finally briefly discuss the odds of finding a macroscopic Klein
bottle hole which is kept stable. It was first argued [18] that
only quantum wormholes, and  hence quantum ringholes and Klein
bottle holes, with nearly the Planck size can be stable, with
larger tunnelings being violently destabilized by quantum effects
produced by catastrophic particle creation taking place near the
chronology horizons. Actually, Hawking even advanced his
chronology protection conjecture [19] for wormholes which can also
be applied to ringholes and Klein bottle holes, preventing the
appearance of closed timelike curves, so making the universe safe
for historians and free of the occurrence of the kind of phenomena
dealt with in this paper. Thus, neither wormholes or ringholes nor
Klein bottle holes could exist due to these quantum fluctuation
instabilities.

However, besides some counter-examples to the Hawking's conjecture
that includes e.g. some compelling argument by Li and Gott [20],
it has been shown [21] that macroscopic wormholes, ringholes and
Klein bottle holes can be stabilized after the cosmic coincidence
time taking place at the onset of the dark energy era by the
accelerating expansion of the universe which induces their throat
to quickly growing comovingly to the super-luminal universal
expansion. On the other hand, similarly to as it happens with
wormholes [22], accretion of phantom energy onto the ringholes and
Klein bottle holes should also induce in them a ultra rapid
swelling up that would circumvent the kind of quantum effects
considered by Hawking so that, such as it also happens with their
above-mentioned size increasing which is comoving to the universal
expansion, the destabilizing quantum effects here cannot act in
time to destroy the tunnel during the current speeding-up of the
universe. Therefore, the odds for all of these tunnelings to exist
and gravitationally act on the light coming from luminous sources
the way we showed before appear to be good enough in the context
of our accelerating universe as for allowing the kind of
interpretation considered in this letter. On the other hand, in
spite of the feature that for ringholes this interpretation just
requires two objects aligned on axis, as opposed to requiring
three objects to be on axis as the interpretation first suggested
in Ref. [14] did, one could likewise think that possibly ringholes
are quite rarer than galaxies so that the former interpretation is
far from being quite more likely than the latter interpretation
either. Thus, at least for the time being, it appears hard to
decide which of these two interpretations should be chosen. Only
more accurate analysis on the involved spectra and on the relative
brightness of the two rings, and mainly the discovery of other
double rings systems, could be used to finally choose which among
these two interpretations is more likely to hold. As to using
Klein bottle holes to check the physical existence of other
universes, it appears just a matter of time to find a double
truncated spiral blurred enough to clearly show a connection with
other universes.

\acknowledgements

\noindent This work was supported by MICINN under research project
no. FIS2008-06332. The author benefited from discussions with C.
Sig\"{u}enza of the {\it Estaci\'{o}n Ecol\'{o}gica de
Biocosmolog\'{\i}a} of Medell\'{\i}n, Spain.

\end{document}